\documentclass{jpsj3}
\usepackage{txfonts}
\usepackage{bm}
\usepackage{braket}
\usepackage{color}

\def\E{\mathrm{e}}
\def\D{\mathrm {d}}

\def\I{\mathrm {i}}
\def\L{\mathrm{L}}
\def\R{\mathrm{R}}
\def\T{\mathrm{T}}

\def\ad{\mathrm{ad}}

\def\ss{\mathrm{s}}

\def\R{\mathrm{R}}
\def\LHS{\mathrm{LHS}}
\def\RHS{\mathrm{RHS}}
\def\SIG{{\bm\sigma}}
\def\hc{\mathrm{H.c.}}

\def\for#1{\quad\mathrm{for\ \ }#1}
\def\abs#1{\left\rvert #1 \right\lvert}

\title{Charge and Spin Transport in Magnetic Tunnel Junctions: Microscopic Theory}

\author{\name{Daisuke \surname{Miura}}\thanks{E-mail address: dmiura@solid.apph.tohoku.ac.jp} and \name{Akimasa \surname{Sakuma}}}
\inst{
\address{Department of Applied Physics, Tohoku University\\Sendai 980-8579}
}
\abst{
We study the charge and spin currents passing through a magnetic tunnel junction (MTJ)
on the basis of a tight-binding model.
The currents are evaluated perturbatively with respect to the tunnel Hamiltonian.
The charge current has the form $A[\bm M_1(t)\times\dot{\bm M}_1(t)]\cdot\bm M_2+B\dot{\bm M}_1(t)\cdot\bm M_2$,
where $\bm M_1(t)$ and $\bm M_2$ denote the directions of the magnetization in the free layer and fixed layer, respectively.
The constant $A$ vanishes when one or both layers are insulators,
{while the constant $B$ disappears when both layers are insulators or the same ferromagnets.}
The first term in the expression for charge current represents dissipation driven by the effective electric field induced by the dynamic magnetization.
In addition, from an investigation of the spin current, we obtain the microscopic expression for the enhanced Gilbert damping constant $\varDelta \alpha$.
We show that $\varDelta\alpha$ is proportional to the tunnel conductance and depends on the bias voltage.
}

\kword{spintronics, magnetic tunnel junction, spin current, spin dynamics}

\begin{document}
\maketitle
\section{Introduction}
Magnetic tunnel junctions (MTJs), which consist of a thin tunnel barrier sandwiched between two ferromagnetic layers \cite{Miyazaki1995L231,PhysRevLett.74.3273,JJAP.43.L588,NatureMater3.862,Parkin},
are promising for their use in magnetic random access memory (MRAM)\cite{839717}.
However, the primary disadvantage of conventional MRAM designs, which employ a current-induced field to write data, is that the writing current increases with the device density. Thus, there has been considerable interest in exploiting spin-transfer torque (STT)\cite{Slonczewski1996L1,PhysRevB.54.9353} instead\cite{albert:3809,JJAP.45.3835,Sun1999157,liu:2871,huai:222510}.
In such an STT MRAM device, the critical current is proportional to the product of the volume and the Gilbert damping constant $\alpha$ of the free layer, making low $\alpha$ an important criterion for electrode materials.

To this end, several studies have explored the dynamics and the distribution of the magnetizations in STT MRAM by using
the Landau--Lifshitz--Gilbert (LLG) equation with an STT term
\cite{miltat:6982,liu:8385,PhysRevB.71.024411,zhang:08G515,zhang:112504}.
However, other torques (spin torques) also act on the dynamic magnetization in the free layer, which form in reaction to the outward flow of spins from the layer: Mizukami {\it et al.} experimentally showed that $\alpha$ increases with the thickness of the nonmagnetic metal (NM) layer in NM/Py/NM films, and that this enhancement continues up to thicknesses of several hundred nanometers \cite{JJAP.40.580}. Their experiment supports the importance of spin torques in the magnetization dynamics of mesoscopic devices such as STT MRAMs. Further, this experimental finding was supported immediately by Tserkovnyak {\it et al.}'s \cite{PhysRevLett.88.117601,PhysRevB.66.224403} theory of spin pumping based on scattering theory, with additional theoretical confirmation by Umetsu {\it et al.} on the basis of the Kubo formula \cite{1742-6596-266-1-012084,umetsu:07D117}.

Several studies have also investigated charge transport in the presence of magnetization dynamics in magnetic multilayers.
It is known that dynamic magnetizations induce an effective electromagnetic field\cite{0022-3719-20-7-003,0305-4470-39-22-022}.
Ohe {\it et al.} simulated the effective electric field induced by the motion of the magnetic vortex core in a magnetic disk\cite{ohe:07C706},
and the field was observed experimentally\cite{ohe:123110}.
Furthermore, Zhang {\it et al.} phenomenologically derived the LLG equation having the STT term induced by this effective electric field\cite{PhysRevLett.102.086601}. And Moriyama {\it et al.} observed the dc voltage across generated by the precession of the magnetization
in an Al/AlO$_x$/Ni$_{80}$Fe$_{20}$/Cu tunnel junction.\cite{PhysRevLett.100.067602}
The origins of this voltage have been discussed from a theoretical standpoint (scattering theory)\cite{PhysRevB.77.180407,PhysRevB.78.020401,PhysRevB.79.054424}.
In addition, charge and spin currents in ferromagnets with magnetizations that slowly vary in space and time have been studied microscopically
\cite{JPSJ.75.113706,JPSJ.77.074701,PhysRevB.81.144405}. These studies employed the s-d model in continuous space and treated the perturbation within the framework of the Keldysh--Green function\cite{Haug,Rammer}.

Similarly, our aim is to describe the charge and spin transport in MTJs in the presence of a voltage across the barrier and the dynamical magnetization in the free layer. This situation just corresponds to an STT MRAM cell during the writing stage.
In this paper, we microscopically describe the charge and spin currents passing through an MTJ.
However, in contrast with previous works that relied on models in continuous space, we calculate the currents on the basis of a tight-binding scheme.
This makes it easier to account for the properties of materials and the space dependence of the magnetization
in magnetic multilayers, such as MTJs, with strongly inhomogeneous magnetic structures.
In the calculations, we consider the voltage and the dynamics of the magnetization in Berry's adiabatic approximation under the assumption that the effective exchange field is larger than the voltage and dynamics.
Our model shows that the charge current induced by the dynamical magnetization has the form
$A[\bm M_\L(t)\times\dot{\bm M}_\L(t)]\cdot\bm M_\R+B\dot{\bm M}_\L(t)\cdot\bm M_\R$,
where $\bm M_\L(t)$ and $\bm M_\R$ denote the directions of the magnetization in the free layer and fixed layer, respectively.
The first term tends to the form given by Tserkovnyak {\it et al.}\cite{PhysRevB.78.020401}, which expressed
the dc current due to the precession of $\bm M_\L(t)$ about $\bm M_\R$ as a special case;
in this sense, our result is a generalization of their work.
Furthermore, from the results concerning spin transport, we successfully derive the enhanced Gilbert damping and propose a microscopic expression for it.

\section{Model and Formalism}
\subsection{Model Hamiltonian}
\begin{figure}
\centering
\includegraphics[scale=0.4]{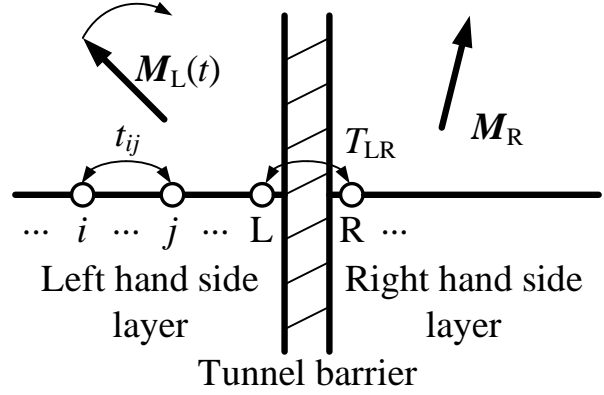}
\caption{{Schematic of one-dimensional magnetic tunnel junction.
$T_{\L\R}$ is the tunneling amplitude and
$t_{ij}$ represents the hopping matrix between sites $i$ and $j$ located
at either side of the interface. $\bm M_\L(t)$ and $\bm M_\R$ denote the directions of the effective exchange fields
for the left (L) and right (R) hand side layer, respectively.}}
\label{fig:fig1}
\end{figure}
We consider the motion of electrons in an effective exchange field.
Furthermore, assume that the ferromagnetic layer on the left-hand side $(\LHS)$ of the MTJ is the free layer; that is, the direction of the field at time $t$ in this layer, $\bm M_\L(t)$, rotates time-dependently {(see Fig. \ref{fig:fig1}).}
Thus, the direction of the field on the right-hand side $(\RHS)$ (fixed layer), $\bm M_\R$, is time-independent.
Note that we ignore the inner structure of the tunnel barrier and account for its properties via the simple tunnel amplitude $T_{\L\R}$ between sites $\L$ and $\R$,
which denote the surfaces on the LHS and RHS, respectively.
In this model, the total Hamiltonian for the MTJ is the sum of
the one dimensional tight-binding Hamiltonians in the ferromagnetic layers,
\begin{align}
\mathcal{H}_\L(t)&:=\sum_{i,j\in \LHS} c_i^\dagger\left[-t_{ij}\hat 1-\delta_{ij}J_\L\bm M_\L(t)\cdot\hat\SIG\right] c_j,\\
\mathcal{H}_\R&:=\sum_{i,j\in \RHS} c_i^\dagger\left[-t_{ij}\hat 1-\delta_{ij}J_\R\bm M_\R\cdot\hat\SIG\right] c_j,
\end{align}
and the tunnel Hamiltonian,
\begin{align}
\mathcal{H}_\T&:=-T_{\L\R}c_\L^\dagger c_\R+\hc,
\end{align}
where
$c_{i\sigma}^\dagger (c_{i\sigma})$ is an operator that creates (annihilates) the $\sigma$ spin electron at site $i$, and
$t_{ij}$ is the hopping integral between sites $i$ and $j$.
The constant $J_\L (J_\R)$ represents the strength of the interaction between the spin of an electron and the effective exchange field on the LHS (RHS) layer;
and $\hat\SIG$ is the Pauli matrix, where hat ` $\hat{}$ ' denotes a $2\times 2$ matrix in spin-space.
\subsection{Adiabatic approximation}
Assuming $J_\L\gg\hbar|\D\bm M_\L(t)/\D t|$, we adopt Berry's adiabatic approximation\cite{Berry} for $\mathcal{H}_\L(t)$:
\begin{align}
c_i(t)&\simeq \hat U_\L(t)\E^{\I\gamma(t)\hat\sigma^z} d_i\for{i\in\LHS},\label{eq:ad}\\
\mathcal{H}_\L(t)&\to \mathcal{H}_\L^\ad:= \sum_{i,j\in \LHS} d_i^\dagger\left[-t_{ij}\hat 1-\delta_{ij}J_\L\hat\sigma^z\right] d_j,
\end{align}
where $c_i(t)$ is in the Heisenberg representation with respect to $\mathcal{H}_\L(t)$, $\hat U_\L(t)$ is a rotation matrix satisfying the equation $\hat U_\L^\dagger(t)\bm M_\L(t)\cdot\hat\SIG \hat U_\L(t)=\hat\sigma^z$, and
$\gamma(t)$ is Berry's phase defined by
\begin{align}
\gamma(t):=\I\int \D t\left[\hat U_\L^\dagger(t)\frac{\D\hat U_\L(t)}{\D t}\right]_{\uparrow\uparrow}\label{eq:berry}.
\end{align}
With the approximation (\ref{eq:ad}), we replace $\mathcal{H}_\T$ with
\begin{align}
\mathcal{H}_\T^\ad(t)&:=-T_{\L\R}d_\L^\dagger\E^{-\I\gamma(t)\hat\sigma^z}\hat U_\L^\dagger(t) \hat U_\R d_\R+\hc,
\end{align}
where $\hat U_\R$ is a rotation matrix satisfying the equation $\hat U_\R^\dagger\bm M_\R\cdot\hat\SIG \hat U_\R=\hat\sigma^z$, and $d_i:=\hat U_\R^\dagger c_i\ \mathrm{for}\ i\in\RHS$.
Finally, our total Hamiltonian is $\mathcal{H}(t):=\mathcal{H}_\L^\ad+\mathcal{H}_\R+\mathcal{H}_\T^\ad(t)$, where
$\mathcal{H}_\R=\sum_{i,j\in \RHS} d_i^\dagger\left[-t_{ij}\hat 1-\delta_{ij}J_\R\hat\sigma^z\right] d_j$.
Thus, a nonequilibrium statistical average of the form $\Braket{d_{i\sigma}(t)d_{i'\sigma'}^\dagger(t')}$ can be derived
perturbatively with respect to $\mathcal{H}_\T^\ad(t)$ using the Keldysh--Green function technique.

\subsection{Charge and spin currents}
The charge current $I^\E(t)$ and spin current $\bm I^\ss(t)$ passing through the MTJ are defined by
\begin{align}
I^\E(t)&:=2\Re \frac{\I}{\hbar}T_{\R\L}\Braket{d_\R^\dagger(t) \hat U_\R^\dagger\hat U_\L(t)\E^{\I\gamma(t)\hat\sigma^z} d_L(t)}\quad[\mathrm{1/s}],\label{eq:Ie}\\
\bm I^\ss(t)&:=2\Re\frac{\I}{\hbar}T_{\R\L}\Braket{d_\R^\dagger(t)\hat U_\R^\dagger\hat\SIG \hat U_\L(t)\E^{\I\gamma(t)\hat\sigma^z}d_L(t)}\quad[\mathrm{1/s}]\label{eq:Is},
\end{align}
where
$\Braket{\cdots}$ denotes a statistical average in $\mathcal{H}(t)$\cite{Haug,Rammer}.

By introducing the lesser function,
\begin{align*}
\left[\hat G^<_{\L\R}(t,t')\right]_{\sigma\sigma'}:=\frac{\I}{\hbar}\Braket{\left[d_\R^\dagger(t') \hat U_\R^\dagger\right]_{\sigma'}\left[\hat U_\L(t)\E^{\I\gamma(t)\hat\sigma^z} d_\L(t)\right]_\sigma},
\end{align*}
eqs. (\ref{eq:Ie}) and (\ref{eq:Is}) can be written in the form
\begin{align}
I^\E(t)&=2\Re T_{\R\L}\tr\hat G^<_{\L\R}(t,t),\\
\bm I^\ss(t)&=2\Re T_{\R\L}\tr\hat\SIG\hat G^<_{\L\R}(t,t).
\end{align}
In the first order in $\mathcal{H}_\T^\ad(t)$, we have
\begin{align}
\hat G^<_{\L\R}(t,t)\simeq -T_{\L\R}\int\D t'
\hat U_\L(t)\E^{\I\gamma(t)\hat\sigma^z}
\hat g_\L(t-t')
\E^{-\I\gamma(t)\hat\sigma^z}
\hat U_\L^\dagger(t)
\hat A(t,t')
\hat U_\R
\hat g_\R(t'-t)
\hat U_\R^\dagger\biggr|^<
,\label{eq:lesser1}
\end{align}
where $<$ denotes the lesser component of Keldysh--Green functions\cite{Haug,Rammer}, and
\begin{align}
\hat A(t,t')&:=\hat U_\L(t)\E^{\I\gamma(t)\hat\sigma^z}\E^{-\I\gamma(t')\hat\sigma^z}\hat U_\L^\dagger(t').
\end{align}
Moreover, we introduce the unperturbed Keldysh--Green functions defined by
\begin{align}
\left[\hat g_\L(t)\right]_{\sigma\sigma'}&:=-\frac{\I}{\hbar}\Braket{\T d_{\L\sigma}(t) d_{\L\sigma'}^\dagger}_0
=-\frac{\I}{\hbar}\Braket{\T d_{\L\sigma}(t) d_{\L\sigma}^\dagger}_0\delta_{\sigma\sigma'}
,\\
\left[\hat g_\R(t)\right]_{\sigma\sigma'}&:=-\frac{\I}{\hbar}\Braket{\T d_{\R\sigma}(t) d_{\R\sigma'}^\dagger}_0
=-\frac{\I}{\hbar}\Braket{\T d_{\R\sigma}(t) d_{\R\sigma}^\dagger}_0\delta_{\sigma\sigma'}
,
\end{align}
where $\T$ is the time-ordering operator on the Keldysh contour, and $\Braket{\cdots}_0$ denotes an equilibrium statistical average in $\mathcal{H}_\L^\ad+\mathcal{H}_\R$.
Since $\hat g_\L(t)$ is the diagonal matrix in spin-space, $\hat g_\L(t)$ and Berry's phase factors commute. Thus, $\hat G^<_{\L\R}(t,t)$ reduces to
\begin{align}
\hat G^<_{\L\R}(t,t)
=-T_{\L\R}
\int\frac{\D E}{2\pi\hbar}
\int\frac{\D E'}{2\pi\hbar}
\E^{-\I E' t/\hbar}
\hat U_\L^\dagger(t)
\hat g_\L(E)
\hat U_\L(t)
\hat A(t,E')
\hat U_\R^\dagger
\hat g_\R(E-E')
\hat U_\R\biggr|^<,\label{eq:lesser2}
\end{align}
where we employ the Fourier transform of a function $f(t)$ with respect to $t$, defined by the relation
\begin{align}
f(E):=\int \D t\E^{\I Et/\hbar} f(t).
\end{align}
$E'$ in eq. (\ref{eq:lesser2}) represents the energy that an electron obtains from the dynamics of the magnetization;
we consider it in the first order:
\begin{align}
\hat G^<_{\L\R}(t,t)\simeq -T_{\L\R}
\int\frac{\D E}{2\pi\hbar}
\hat U_\L^\dagger(t)
\hat g_\L(E)
\hat U_\L(t)
\hat U_\R^\dagger
\hat g_\R(E)
\hat U_\R
\biggr|^<\notag\\
-T_{\L\R}
\int\frac{\D E}{2\pi\hbar}
\hat U_\L^\dagger(t)
\hat g_\L(E)
\hat U_\L(t)
\frac{\hbar}{\I}
\left.\frac{\D \hat A(t,t')}{\D t'}\right|_{t'=t}
\hat U_\R^\dagger
\frac{\D\hat g_\R(E)}{\D E}
\hat U_\R
\biggr|^<.
\end{align}
Then, using the relations
\begin{align}
\frac{\hbar}{\I}
\frac{\D \hat A(t,t')}{\D t'}\biggr|_{t'=t}
&=
\frac{\hbar}{2}\hat\SIG\cdot\bm M_\L(t)\times\frac{\D \bm M_\L(t)}{\D t},\\
\hat U_\L^\dagger(t)
\hat g_\L(E)
\hat U_\L(t)
&=
\hat 1 \bar g_\L(E)+\hat\SIG\cdot\bm M_\L(t)\varDelta g_\L(E),\\
\hat U_\R^\dagger
\hat g_\R(E)
\hat U_\R
&=
\hat 1 \bar g_\R(E)+\hat\SIG\cdot\bm M_\R\varDelta g_\R(E),
\end{align}
where
\begin{align}
\bar g_{\L (\R)}(E)&:=\frac{1}{2}\tr\hat g_{\L (\R)}(E)=\frac{\left[\hat g_{\L (\R)}(E)\right]_{\uparrow\uparrow}
+\left[\hat g_{\L (\R)}(E)\right]_{\downarrow\downarrow}
}{2},\\
\varDelta g_{\L (\R)}(E)&:=\frac{1}{2}\tr\hat\sigma^z\hat g_{\L (\R)}(E)
=\frac{\left[\hat g_{\L (\R)}(E)\right]_{\uparrow\uparrow}
-\left[\hat g_{\L (\R)}(E)\right]_{\downarrow\downarrow}
}{2}
,
\end{align}
we can decompose $\hat G^<_{\L\R}(t,t)$ into two terms:
\begin{align}
\hat G^<_{\L\R}(t,t)=\hat 1 G^<_{\L\R}(t,t)+\hat\SIG\cdot \bm G^<_{\L\R}(t,t).
\end{align}
Here we define
\begin{align}
G^<_{\L\R}(t,t)&:=
-T_{\L\R}\int\frac{\D E}{2\pi\hbar}\Biggl[
\bar g_\L(E)\bar g_\R(E)+\bm M_\L(t)\cdot\bm M_\R\varDelta g_\L(E)\varDelta g_\R(E)\notag\\
&+\frac{\hbar}{2}\bm M_\L(t)\times\frac{\D\bm M_\L(t)}{\D t}\cdot\bm M_\R\bar g_\L(E)\frac{\D\varDelta g_\R(E)}{\D E}
-\I\frac{\hbar}{2}\frac{\D\bm M_\L(t)}{\D t}\cdot\bm M_\R\varDelta g_\L(E)\frac{\D\varDelta g_\R(E)}{\D E}
\Biggr]^<,\\
\bm G^<_{\L\R}(t,t)&:=
-T_{\L\R}\int\frac{\D E}{2\pi\hbar}\Biggl[
\bm M_\L(t)\varDelta g_\L(E)\bar g_\R(E)+\bm M_\R\bar g_\L(E)\varDelta g_\R(E)
+\I\bm M_\L(t)\times\bm M_\R\varDelta g_\L(E)\varDelta g_\R(E)\notag\\
&+\frac{\hbar}{2}\bm M_\L(t)\times\frac{\D\bm M_\L(t)}{\D t}\bar g_\L(E)\frac{\D \bar g_\R(E)}{\D E}
-
\I\frac{\hbar}{2}\frac{\D\bm M_\L(t)}{\D t}\varDelta g_\L(E)\frac{\D \bar g_\R(E)}{\D E}
\notag\\
&+
\I\frac{\hbar}{2}\left\{\bm M_\L(t)\times\frac{\D\bm M_\L(t)}{\D t}\right\}\times\bm M_\R\bar g_\L(E)\frac{\D \varDelta g_\R(E)}{\D E}
+
\frac{\hbar}{2}\frac{\D\bm M_\L(t)}{\D t}\times\bm M_\R\varDelta g_\L(E)\frac{\D \varDelta g_\R(E)}{\D E}
\Biggr]^<.
\end{align}
$I^\E(t)$ and $\bm I^\ss(T)$ are expressed in terms of $G^<_{\L\R}(t,t)$ and $\bm G^<_{\L\R}(t,t)$ as follows:
\begin{align}
I^\E(t)&=4\Re T_{\R\L} G^<_{\L\R}(t,t),\\
\bm I^\ss(t)&=4\Re T_{\R\L}\bm G^<_{\L\R}(t,t).
\end{align}
Finally, taking the lesser components, we obtain the following in the low-temperature limit:
\begin{align}
I^\E(t)
&=
2\pi
\abs{T_{\L\R}}^2\Biggl\{
\bar \rho_\L(\mu)\varDelta \rho_\R(\mu)
\bm M_\L(t)\times\frac{\D\bm M_\L(t)}{\D t}\cdot\bm M_\R\notag\\
&-
\int^\mu\D E \Biggl[\varDelta \rho_\L(E)\frac{\D \varDelta \chi_\R(E)}{\D E}-
\frac{\D\varDelta \chi_\L(E)}{\D E}\varDelta \rho_\R(E)
\Biggr]
\frac{\D\bm M_\L(t)}{\D t}\cdot\bm M_\R
\Biggr\},\label{eq:chargecurrent}\\
\bm I^\ss(t)
&=
\frac{4\pi\abs{T_{\L\R}}^2}{\hbar}
\int^\mu\D E
\bigl[
\varDelta \rho_\L(E) \varDelta \chi_\R(E)
+
\varDelta \chi_\L(E) \varDelta \rho_\R(E)
\bigr]
\bm M_\L(t)\times\bm M_\R
\notag\\
&+
\frac{4\pi\abs{T_{\L\R}}^2}{\hbar}
\Biggl\{
\bar \rho_\L(\mu)\bar \rho_\R(\mu)\frac{\hbar}{2}\bm M_\L(t)\times\frac{\D\bm M_\L(t)}{\D t}\notag\\
&-\int^\mu\D E\Biggl[\varDelta \rho_\L(E)\frac{\D \bar \chi_\R(E)}{\D E}-\frac{\D\varDelta \chi_\L(E)}{\D E}\bar \rho_\R(E)\Biggr]\frac{\hbar}{2}\frac{\D\bm M_\L(t)}{\D t}\notag\\
&+\int^\mu\D E\Biggl[\bar \rho_\L(E)\frac{\D \varDelta \chi_\R(E)}{\D E}-\frac{\D \bar \chi_\L(E)}{\D E}\varDelta \rho_\R(E)\Biggr]\frac{\hbar}{2}\Biggl[\bm M_\L(t)\times\frac{\D\bm M_\L(t)}{\D t}\Biggr]\times\bm M_\R
\notag\\
&+\varDelta \rho_\L(\mu)\varDelta \rho_\R(\mu)\frac{\hbar}{2}\frac{\D\bm M_\L(t)}{\D t}\times\bm M_\R\Biggr\}\label{eq:spincurrent},
\end{align}
Here $\mu$ is the chemical potential of the system. $\bar\rho_{\L (\R)}(E)$ and $\varDelta\rho_{\L (\R)}(E)$ are the spin-averaged local density of states (DOS) and the spin polarization of the local DOS, respectively, at the LHS (RHS) layer surface, defined by
\begin{align}
\bar\rho_{\L (\R)}(E)&:=-\frac{1}{\pi}\Im\bar g^\mathrm{r}_{\L (\R)}(E),\\
\varDelta\rho_{\L (\R)}(E)&:=-\frac{1}{\pi}\Im\varDelta g^\mathrm{r}_{\L (\R)}(E),
\end{align}
where $g^\mathrm{r}$'s are retarded Green's functions from the calculations taking the lesser component.
Furthermore, the $\chi$'s are defined as the real parts of the retarded Green's functions,
\begin{align}
\bar\chi_{\L (\R)}(E)&:=\frac{1}{\pi}\Re\bar g^\mathrm{r}_{\L (\R)}(E),\\
\varDelta\chi_{\L (\R)}(E)&:=\frac{1}{\pi}\Re\varDelta g^\mathrm{r}_{\L (\R)}(E).
\end{align}
\section{Discussion and Summary}
\subsection{Charge current}
The form $A[\bm M_\L(t)\times\dot{\bm M}_\L(t)]\cdot\bm M_\R+B\dot{\bm M}_\L(t)\cdot\bm M_\R$ of the charge current
{(eq. (\ref{eq:chargecurrent}))
driven by the magnetization dynamics
is consistent with previous works;
The first term tends to the form given by Tserkovnyak {\it et al.}\cite{PhysRevB.78.020401} in the special case where $\bm M_\L(t)$ precesses around $\bm M_\R$, as discussed in \S\ref{sec:EFA}.
And Xiao {\it et al.}\cite{PhysRevB.77.180407} have derived the same form for the charge current passing through the MTJ on the basis of scattering theory in the continuum space,
whereas, we calculate the current on the basis of the tight-binding model.
New insights which we found out in this study are as follows.}
If the electronic structure in the two layers of the MTJ is the same [i.e., $\varDelta\rho_\L(E)=\varDelta\rho_\R(E)$ and $\varDelta\chi_\L(E)=\varDelta\chi_\R(E)$]
or
both layers are insulators,
we have $B=0$.
However, if either layer is metallic, finite $B$ should be measured because the real part of the retarded Green's function remains finite, which reflects virtual transitions through forbidden bands.
The term $[\bm M_\L(t)\times\dot{\bm M}_\L(t)]\cdot\bm M_\R$ in eq. (\ref{eq:chargecurrent}) represents the charge current driven by the effective electric field (i.e., the spin electric field), as mentioned in \S\ref{sec:EFA}.
In other words, the effective electrochemical potential of the free layer is changed by the dynamics of $\bm M_\L(t)$, and
the resultant difference in electrochemical potentials between the two layers manifests as a bias voltage\cite{PhysRevB.78.020401}.
This situation may be realized when a barrier exists between the electrode and lead,
or when the diffusion constant of the free layer is small enough to maintain the changed effective chemical potential.
Otherwise, this charge current will flow back to the reservoir connected to the free layer without tunneling through the barrier of the MTJ.

\subsection{Spin current}
The term $\bm M_\L(t)\times\bm M_\R$ in eq. (\ref{eq:spincurrent}) represents the static effective Heisenberg coupling between $\bm M_\L(t)$ and $\bm M_\R$.
That is, the equation of motion for $\bm M_\L(t)$ described by this spin current corresponds to the equation $\frac{\D\bm M_\L(t)}{\D t}=\frac{J_\mathrm{eff}}{\hbar |\bm S_\L(t)|}\bm M_\L(t)\times\bm M_\R$ [$\bm S_\L(t)$ is defined by eq. (\ref{eq:SL})].
This affords a Heisenberg coupling energy of $-J_\mathrm{eff}\bm M_\L(t)\cdot\bm M_\R$,
where
\begin{align}
J_\mathrm{eff}&:=-4\pi|T_{\L\R}|^2\int^\mu\D E
\left[
\varDelta \rho_\L(E) \varDelta \chi_\R(E)
+
\varDelta \chi_\L(E) \varDelta \rho_\R(E)
\right]\\
&=
\frac{1}{\pi}\int^\mu\D E\Im G^{\mathrm{r}\uparrow}_{\L\R}(E)\Delta_\R (E) G_{\R\L}^{\mathrm{r}\downarrow}(E)\Delta_\L(E),\\
G^{\mathrm{r}\sigma}_{ij}(E)&:=g_{i\sigma}^\mathrm{r}(E)T_{ij}g_{j\sigma}^\mathrm{r}(E),\\
\Delta_i(E)&:=g_{i\uparrow}^\mathrm{r}(E)^{-1}-g_{i\downarrow}^\mathrm{r}(E)^{-1}.
\end{align}
$\Delta_i(E)$ describes the exchange splitting at site $i$, and this result agrees with the expression presented by Liechtenstein {\it et al.}\cite{Liechtenstein198765}

Let us consider the term $\bm M_\L\times\frac{\D\bm M_\L(t)}{\D t}$ in eq. (\ref{eq:spincurrent}):
\begin{align}
2\pi|T_{\L\R}|^2\bar\rho_\L(\mu) \bar\rho_\R(\mu)\bm M_\L(t)\times\frac{\D\bm M_\L(t)}{\D t}
=
\frac{\hbar}{2e^2}\bar\Gamma\bm M_\L(t)\times\frac{\D\bm M_\L(t)}{\D t},
\end{align}
where {$e>0$ is the elementary charge and} $\bar\Gamma$ is the tunnel conductance of the MTJ,
\begin{align}
\bar\Gamma:=\frac{4\pi|T_{\L\R}|^2e^2}{\hbar}\bar\rho_\L(\mu) \bar\rho_\R(\mu).
\end{align}
This term describes the spin pumping in the MTJ and affords the following microscopic expression for the enhanced Gilbert damping constant:
\begin{align}
\varDelta\alpha=\frac{\hbar}{2e^2}\frac{\bar\Gamma}{|\bm S_\L(t)|}\label{eq:dalpha},
\end{align}
where $\bm S_\L(t)$ is the total spin polarization of the electrons in the LHS layer,
\begin{align}
\bm S_\L(t):=2\sum_{i\in\LHS}\int^\mu\D E\varDelta\rho_i(E)\bm M_\L(t)\label{eq:SL}.
\end{align}
Equation (\ref{eq:dalpha}) agrees with the corrected Gilbert damping constant derived by Zhang {\it et al.}\cite{PhysRevLett.102.086601} phenomenologically after considering the effect of the spin electric field induced by the dynamic magnetization.
In addition, in the present formulation, from the fact that $\varDelta\alpha$ vanishes if one ignores Berry's phase (\ref{eq:berry})\cite{miura:07C909}, it follows that one of the origins of spin pumping is the spin electric field.
As a consequence of this, $\varDelta\alpha$ is proportional to the conductance $\bar\Gamma$.

The size dependence of $\varDelta\alpha$ can be described as follows:
\begin{align}
\varDelta\alpha\propto \frac{1}{{\lambda}},
\end{align}
{where $\lambda$ is thickness of the free layer,} because $|\bm S_\L(t)|$ is roughly proportional to the volume of the free layer, and $\bar \Gamma$ to the cross-sectional area of the barrier.
\subsection{Analysis of effective field}\label{sec:EFA}
For a more transparent physical interpretation of the currents, we rewrite eqs. (\ref{eq:chargecurrent}) and (\ref{eq:spincurrent}) as follows:
\begin{align}
-e I^\E(t)&=\sum_{\sigma=\pm 1}\left[
\Gamma_\sigma^\R\bm\varepsilon_\sigma^1(t)+\gamma_\sigma^\L\bm\varepsilon_\sigma^2(t)
\right]\cdot\bm M_\R,\label{eq:chargecurrent2}\\
-e\bm I^\ss(t)&=\left[\frac{eJ_\mathrm{eff}}{\hbar}\bm M_\L(t)-\varDelta\Gamma\frac{\hbar}{2e}\frac{\D\bm M_\L(t)}{\D t}\right]\times\bm M_\R\notag\\
&+
\sum_{\sigma=\pm 1}\sigma\Bigl\{
\Gamma_\sigma^\R\bm \varepsilon_\sigma^1(t)
+
\Bigl[
\gamma_\sigma^\L
\bm\varepsilon_\sigma^2(t)
\cdot
\bm M_\R
\Bigr]
\bm M_\L(t)
-
\Bigl[
\gamma_\sigma^\R
+\bm M_\L(t)\cdot\bm M_\R\gamma_\sigma^\L
\Bigr]
\bm \varepsilon_\sigma^2(t)
\Bigr\},\label{eq:spincurrent2}
\end{align}
where the ``conductances'' are defined by
\begin{align}
\Gamma_\sigma^\R&:=\frac{2\pi|T_{\L\R}|^2e^2\bar\rho_\L(\mu)\rho_{\R\sigma}(\mu)}{\hbar},\\
\varDelta\Gamma&:=\frac{4\pi|T_{\L\R}|^2e^2\varDelta\rho_\L(\mu)\varDelta\rho_{\R}(\mu)}{\hbar},\\
\gamma_\sigma^{\L}&:=-\frac{2\pi|T_{\L\R}|^2e^2}{\hbar}\int^\mu\D E\left[\rho_{\L\sigma}(E)\frac{\D\varDelta\chi_\R (E)}{\D E}-\frac{\D\chi_{\L\sigma}(E)}{\D E}\varDelta\rho_{\R}(E)\right],\\
\gamma_\sigma^{\R}&:=-\frac{2\pi|T_{\L\R}|^2e^2}{\hbar}\int^\mu\D E\left[\rho_{\R\sigma}(E)\frac{\D\varDelta\chi_\L (E)}{\D E}-\frac{\D\chi_{\R\sigma}(E)}{\D E}\varDelta\rho_{\L}(E)\right],
\end{align}
and the effective driving fields can be defined by
\begin{align}
\bm \varepsilon^1_\sigma(t)&:=-\frac{\sigma\hbar}{2e}\bm M_\L(t)\times\frac{\D\bm M_\L(t)}{\D t},\\
\bm \varepsilon^2_\sigma(t)&:=-\frac{\sigma\hbar}{2e}\frac{\D\bm M_\L(t)}{\D t}.
\end{align}
The conductances represented by a capital letter denote the ``Fermi surface terms,'' whereas those represented by a small letter denote the ``Fermi sea terms.''
The spin-dependent effective voltage $\bm\varepsilon^1_\sigma(t)\cdot\bm M_\R$ in eq. (\ref{eq:chargecurrent2}) just corresponds to the spin electric field between the layers.
To compare the expressions obtained in continuous space and in discrete space, let us define the correspondences
$\bm M(\bm r,t):=\bm M_\L(t)$ and $\bm M(\bm r+\varDelta\bm r,t):=\bm M_\R$,
where $\varDelta\bm r$ denotes the barrier thickness.
Then we find
$\bm\varepsilon^1_\sigma(t)\cdot\bm M_\R\simeq\varDelta r^i \left(-\frac{\sigma\hbar}{2e}\right)\frac{\partial\bm M(\bm r,t)}{\partial t}\times\frac{\partial\bm M(\bm r,t)}{\partial x^i}\cdot\bm M(\bm r,t)$, which is well-known as the spin electric field.
When $\bm M_\L(t)$ steadily precess about the direction of $\bm M_\R$ with a constant cone angle $\theta$ and a constant frequency $\omega$,
the voltage is time-independent:
\begin{align}
\bm\varepsilon^1_\sigma(t)\cdot\bm M_\R=-\sigma\frac{\hbar\omega}{2e}\sin^2\theta,
\end{align}
This affords an estimate $\hbar\omega/2e\sim 20$ $\mu$V at $10$ GHz. The Fermi sea term in eq. (\ref{eq:chargecurrent2}) vanishes in this case.
This result is in good agreement with that of {Xiao {\it et al.}\cite{PhysRevB.77.180407} and} Tserkovnyak {\it et al.}\cite{PhysRevB.78.020401}
Note that in general the Fermi sea term is certainly the ac current.

Next, let us consider the spin current (\ref{eq:spincurrent2}).
The terms including $\Gamma_\sigma^\R\bm \varepsilon_\sigma^1(t)+\Bigl[\gamma_\sigma^\L\bm\varepsilon_\sigma^2(t)\cdot\bm M_\R\Bigr]\bm M_\L(t)$
describe the spin transport due to the spin $\sigma$ component of the charge current.
By considering $\bm\varepsilon^2_\sigma(t)$ as a driving force,
we can interpret the term $\left[
\gamma_\sigma^\R
+\bm M_\L(t)\cdot\bm M_\R\gamma_\sigma^\L
\right]
\bm \varepsilon_\sigma^2(t)$ as the ``tunneling magnetoresistance (TMR) effect'' in spin transport.
\subsection{Effects of bias voltage}
Finally, we consider the charge and spin transport in the presence of a bias voltage $V(t)$ across the MTJ.
In Berry's adiabatic approximation under the assumption $J_\L\gg e|V(t)|$,
the effects of $V(t)$ can be included by replacing eq. (\ref{eq:ad}) with
\begin{align}
c_i(t)&\simeq \E^{-\frac{\I e}{\hbar}\int\D t V(t)}\hat U_\L(t)\E^{\I\gamma(t)\hat\sigma^z} d_i\for{i\in\LHS}.
\end{align}
In the first order in $\frac{\D V(t)}{\D t}$,
the effective exchange constant and conductances differ as follows:
\begin{align}
J_\mathrm{eff}&\to J_\mathrm{eff}+(\gamma_\uparrow^\L-\gamma_\downarrow^\L)\frac{\hbar}{e}V(t)+\varDelta\Gamma\frac{\hbar^2}{2e}\frac{\D}{\D\mu}\ln\left[\frac{\varDelta\rho_\L(\mu)}{\varDelta\rho_\R(\mu)}\right]\frac{\D V(t)}{\D t},\\
\Gamma_\sigma^\R&\to\Gamma_\sigma^\R
\left\{
1-\frac{\D}{\D\mu}\ln\left[\frac{\bar\rho_\L(\mu)}{\rho_{\R\sigma}(\mu)}\right]eV(t)
-\int^\mu\D E\frac{\bar\rho_\L(E)\frac{\D^3\chi_{\R\sigma}(E)}{\D E^3}
-
\frac{\D^3\bar\chi_{{\L}}(E)}{\D E^3}\rho_{{\R}\sigma}(E)
}{\bar\rho_\L(\mu)\rho_{\R\sigma}(\mu)}
\frac{e\hbar}{2}\frac{\D V(t)}{\D t}
\right\},\\
\varDelta\Gamma&\to\varDelta\Gamma
\left\{1
-\frac{\D}{\D\mu}\ln\left[\frac{\varDelta\rho_\L(\mu)}{\varDelta\rho_\R(\mu)}\right]eV(t)
-\int^\mu\D E\frac{\varDelta\rho_\L(E)\frac{\D^3\varDelta\chi_{\R}(E)}{\D E^3}
-
\frac{\D^3\varDelta\chi_{{\L}}(E)}{\D E^3}\varDelta\rho_{{\R}}(E)
}{\varDelta\rho_\L(\mu)\varDelta\rho_\R(\mu)}
\frac{e\hbar}{2}\frac{\D V(t)}{\D t},
\right\}
\\
\gamma_\sigma^\L&\to\gamma_\sigma^\L-\frac{2\pi|T_{\L\R}|^2 e^2}{\hbar}\int^\mu\D E\left[\rho_{\L\sigma}(E)\frac{\D^2\varDelta\chi_\R(E)}{\D E^2}+\frac{\D^2\chi_{\L\sigma}(E)}{\D E^2}\varDelta\rho_\R(E)\right]eV(t)
\notag\\
&+\frac{2\pi|T_{\L\R}|^2e^2}{\hbar}
\left[
\frac{\D\rho_{\L\sigma}(\mu)}{\D \mu}\frac{\D\varDelta\rho_{\R}(\mu)}{\D \mu}
-
\frac{\D^2\rho_{\L\sigma}(\mu)}{\D \mu^2}\varDelta\rho_{\R}(\mu)
-
\rho_{\L\sigma}(\mu)\frac{\D^2\varDelta\rho_{\R}(\mu)}{\D \mu^2}
\right]\frac{e\hbar}{2}\frac{\D V(t)}{\D t},\\
\gamma_\sigma^\R&\to\gamma_\sigma^\R
-\frac{2\pi|T_{\L\R}|^2 e^2}{\hbar}\int^\mu\D E\left[\rho_{\R\sigma}(E)\frac{\D^2\varDelta\chi_\L(E)}{\D E^2}+\frac{\D^2\chi_{\R\sigma}(E)}{\D E^2}\varDelta\rho_\L(E)\right]eV(t)
\notag\\
&+\frac{2\pi|T_{\L\R}|^2e^2}{\hbar}
\left[
\frac{\D\rho_{\R\sigma}(\mu)}{\D \mu}\frac{\D\varDelta\rho_{\L}(\mu)}{\D \mu}
-
\frac{\D^2\rho_{\R\sigma}(\mu)}{\D \mu^2}\varDelta\rho_{\L}(\mu)
-
\rho_{\R\sigma}(\mu)\frac{\D^2\varDelta\rho_{\L}(\mu)}{\D \mu^2}
\right]\frac{e\hbar}{2}\frac{\D V(t)}{\D t}.
\end{align}
\normalsize
In addition,
a term describing the TMR effect,
\begin{align}
&\frac{1}{e}\left[\bar\Gamma
+\varDelta\Gamma\bm M_\L(t)\cdot\bm M_\R
\right] V(t)
+
\frac{1}{-e}\left[
\bar\gamma
+
\varDelta\gamma
\bm M_\L(t)\cdot\bm M_\R
\right]\frac{\hbar}{2}\frac{\D V(t)}{\D t}
\end{align}
appears in the charge current, where
\begin{align*}
\bar\gamma&:=\frac{4\pi|T_{\L\R}|^2e^2}{\hbar}\int^\mu \D E\left[
\bar\rho_\L(E)\frac{\D^2\bar\chi_\R(E)}{\D E^2}
+
\frac{\D^2 \bar\chi_\L(E)}{\D E^2}\bar\rho_\R(E)\right],
\\
\varDelta\gamma&:=
\frac{4\pi|T_{\L\R}|^2e^2}{\hbar}
\int^\mu \D E\left[
\varDelta\rho_\L(E)\frac{\D^2\varDelta\chi_\R(E)}{\D E^2}
+
\frac{\D^2 \varDelta\chi_\L(E)}{\D E^2}\varDelta\rho_\R(E)
\right].
\end{align*}
For the spin current, a term describing the STT effect,
\begin{align*}
\frac{1}{e}\left[(\Gamma_\uparrow^\L-\Gamma_\downarrow^\L)V(t)-(\gamma_\uparrow^\R+\gamma_\downarrow^\R)\frac{\hbar}{2e}\frac{\D V(t)}{\D t}\right]\bm M_\L(t)
+
\frac{1}{e}\left[(\Gamma_\uparrow^\R-\Gamma_\downarrow^\R)V(t)+(\gamma_\uparrow^\L+\gamma_\downarrow^\L)\frac{\hbar}{2e}\frac{\D V(t)}{\D t}\right]\bm M_\R
\end{align*}
is added, where
\begin{align}
\Gamma_\sigma^\L&:=\frac{2\pi|T_{\L\R}|^2e^2\rho_{\L\sigma}(\mu)\bar\rho_\R(\mu)}{\hbar}.
\end{align}
Then for the Gilbert damping,
since $\bar\Gamma=\Gamma_\uparrow^\R+\Gamma_\downarrow^\R$,
$\varDelta\alpha$ changes as follows:
\begin{align}
\varDelta\alpha&\to
\varDelta\alpha\left[1-\frac{\D}{\D\mu}\ln\left\{\frac{\bar\rho_\L(\mu)}{\bar\rho_\R(\mu)}\right\} eV(t)
-\int^\mu\D E\frac{
\bar\rho_\L(E)\frac{\D^3\bar\chi_{\R}(E)}{\D E^3}
-
\frac{\D^3\bar\chi_{{\L}}(E)}{\D E^3}\bar\rho_{{\R}}(E)
}{
\bar\rho_\L(\mu)\bar\rho_\R(\mu)
}
\frac{e\hbar}{2}\frac{\D V(t)}{\D t}
\right].
\end{align}
This result indicates that when writing data to an STT MRAM cell, the damping of the magnetization dynamics is influenced by not only the spin pumping but also the bias voltage.
However, the effect of the bias voltage on $\varDelta\alpha$ vanishes when both electrodes have the same electronic structure.

In summary, we derived, at the microscopic level, the charge and spin currents passing through an MTJ in response to arbitrary motion of the magnetization in the free layer.
The charge current consists of both Fermi surface and Fermi sea terms.
The Fermi surface term is driven by the spin electric field and manifests as a dc current for steady precession of $\bm M_\L(t)$ in the direction of $\bm M_\R$,
whereas the Fermi sea term is due to virtual transitions and essentially manifests as the ac current.
With regard to spin transport, we focused particularly on the enhanced Gilbert damping (or the spin pumping effect) and
thus obtained the microscopic expression for the enhanced Gilbert damping constant $\varDelta\alpha=\frac{\hbar}{2e^2}\frac{\bar\Gamma}{|\bm S_\L(t)|}$.
Under a bias voltage, the DOSs of the two layers in the MTJ are shifted.
Thus, the bias voltage changes the effective exchange constant and the conductances,
thus producing modulation of $\varDelta\alpha$. All the conductances consist of the tunneling amplitude $T_{\L\R}$ and the local DOS on the surfaces of the layers;
the real part of a retarded Green's function can be obtained from the imaginary part (namely, the local DOS) via the Kramers--Kronig relationship.
In this formulation, the properties of the barrier layer material are considered in the local DOS, which can be easily obtained by first-principles calculations.
\bibliographystyle{jpsj}

\end{document}